\documentclass[twocolumn,aps,prl,groupedaddress]{revtex4}
\usepackage{graphicx}
\usepackage{color}
\usepackage{amsmath,amssymb}
\usepackage{marvosym}
\usepackage{soul}

\newcommand{\be}{\begin{equation}}
\newcommand{\ee}{\end{equation}}
\newcommand{\bea}{\begin{eqnarray}}
\newcommand{\eea}{\end{eqnarray}}

\newcommand{\lb}{\left[}
\newcommand{\rb}{\right]}
\newcommand{\lp}{\left(}
\newcommand{\rp}{\right)}

\newcommand{\gammatr}{\gamma_{\rm tr}}
\newcommand{\tautr}{\tau_{\rm tr}}
\renewcommand{\Im}{{\rm Im}\,}
\renewcommand{\Re}{{\rm Re}\,}
\renewcommand{\vec}[1]{{\bf #1}}
\def\zerop{\big|_{0^+}}
\def\zerom{\big|_{0^-}}

\def\zerox{\big|_{x=0}}

\def\invnm{\ \rm nm^{-1}}
\def\unitnm{\ \rm nm}
\def\unitev{\ \rm eV}
\def\unitmev{\ \rm meV}
\def\unitfs{\ \rm fs}
\def\unitps{\ \rm ps}

\def\unitmobility{\ \rm cm^2/V s}
\renewcommand{\tilde}[1]{{\widetilde #1}}
\def\nn{\nonumber\\}

\newcommand{\bib}[5]{ {#1} {\emph{#2}} {\bf{#5}}, {#3}, {#4}.}
\begin{document}

\title{Long-lived domain wall plasmons in gapped bilayer graphene} 
\author{Eddwi H. Hasdeo$^1$ and Justin C. W. Song$^{1,2}$}
\affiliation{$^1$Institute of High Performance Computing, Agency for Science,
Technology, and Research, Singapore 138632\\ 
$^2$Division of Physics and Applied Physics, Nanyang Technological University, Singapore 637371}
\begin{abstract}Topological domain walls in dual-gated gapped bilayer
    graphene host edge states that are gate-tunable and valley
    polarized. 
    Here we predict that plasmonic collective modes can propagate
    along these topological domain walls even at zero bulk density,
    and possess a markedly different character from that of bulk
    plasmons. Strikingly, domain wall plasmons are extremely
    long-lived, with plasmon lifetimes that can be orders of magnitude
    larger than the transport scattering time in the bulk. While most
    pronounced at low temperatures, long domain wall plasmon lifetimes
    persist even at room temperature with values up to a few
    picoseconds. Domain wall plasmons possess a rich phenomenology
    including a wide range of frequencies (up to the
    mid-infrared), tunable sub-wavelength electro-magnetic confinement
    lengths, as well as a valley polarization for forward/backward
    propagating modes. Its unusual features render them a new tool for
    realizing low-dissipation plasmonics that transcend the
    restrictions of the bulk.
\end{abstract} 
\maketitle

Edge states are a hallmark of the peculiar twisting of crystal
wavefunctions in topological
materials~\cite{kane05,bernevig06,konig07}, and host a fermiology that
departs from that of its parent bulk
\cite{murakami06,bernevig06a,hughes08}. Domain wall edge states (DWS)
in gapped bilayer graphene are a particularly interesting
example. Arising when the sign of the local gap in gapped bilayer
graphene flips in real
space~\cite{martin08,jeiljung11,fanzhang13,eunah13,fengwang14,helin16,
  junzhu16}, DWS manifest in a number of
different settings, e.g., at stacking faults (AB- and
BA-)~\cite{fanzhang13,eunah13,fengwang14,helin16} or in a split
dual-gate geometry wherein perpendicular applied electric field in
adjacent regions have opposite
signs~\cite{martin08,fanzhang13,junzhu16}.  Domain walls host counter-propagating one dimensional (1D) edge states
(DWS) living in separate $K$ and $K'$ valleys~\cite{martin08,fanzhang13,eunah13}, with valley filtered
currents that are robust to disorder~\cite{jeiljung11}. In contrast to
helical edge states in intrinsic topological insulators
\cite{kane05,bernevig06,konig07}, DWS in gapped bilayer graphene enjoy
large and tunable bulk gaps up to $200\, {\rm meV}$~\cite{castro07}
allowing their unusual behavior to manifest even at room temperature.

\begin{figure*}
  \center
\includegraphics[width=12cm]{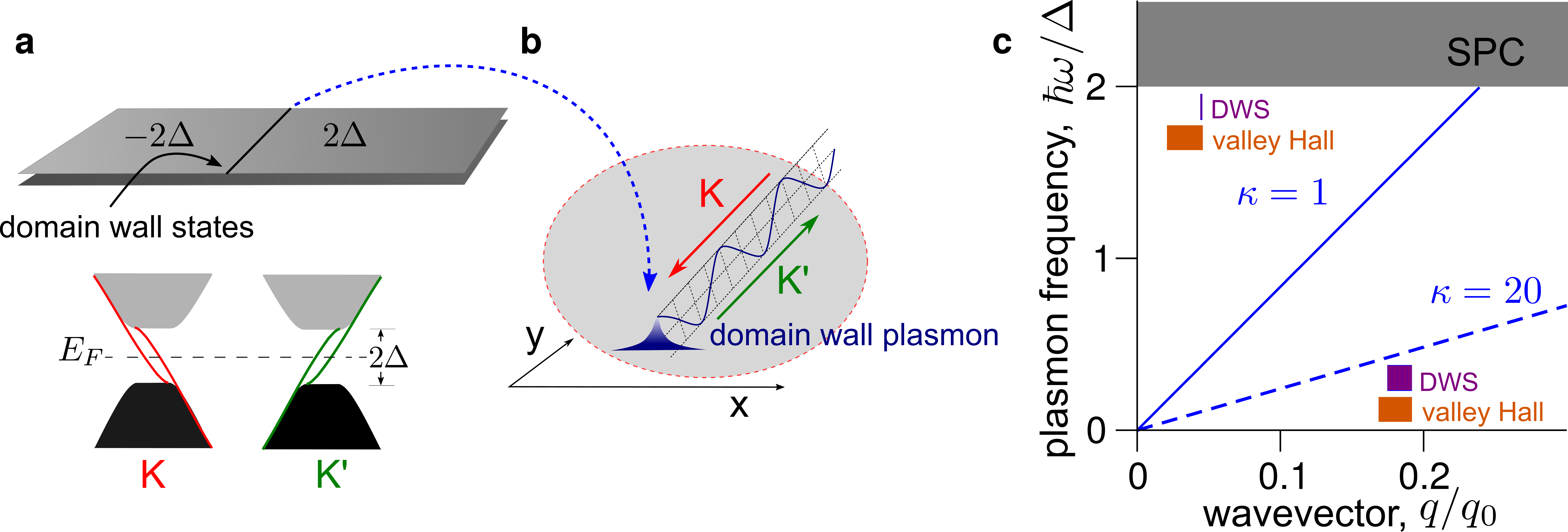}
\caption{\label{Fig1}{\bf a} Domain wall edge states (DWS) localized
  at $x=0$ emerge when the effective band gap of bilayer
    graphene (see text) in adjacent regions have opposite signs:
    $-2\Delta$ on the left, and $+2\Delta$ and the right. DWS are
    valley-helical states located inside the bulk band gap:
    backward/forward propagating
    correspond to $K$ (red lines) and $K'$ (green lines). {\bf b}
    Collective modes of carriers in the domain wall edge states manifest as DWPs,
    which are propagating charge density waves.  DWP current
    at $K$ ($K'$) valley predominantly propagates along the $-y$
    ($+y$) direction. {\bf c} DWP dispersion for $\kappa=1$ (solid
    line) and $\kappa=20$ (dashed line), see
    Eq.~(\ref{eq:plasdisp}). Purple and orange bars show contributions
    from edge states (DWS) and bulk undergap valley Hall motion
    respectively. Shaded region at $\hbar\omega/\Delta\ge 2$
  indicates the single particle continuum (SPC). Parameter values
  used: $ \sigma_H/v_0=1.5$, $\Delta=0.1$~eV and
  $q_0=\sqrt{\Delta\gamma_1}/\hbar v_{F}=0.26\invnm$. }
\end{figure*}

Here we show that the collective motion of carriers in DWS manifest
unusual plasmon modes --- domain wall edge plasmons (DWPs) --- whose
characteristics are distinct from conventional bulk plasmons
(Fig.~\ref{Fig1}). Arising from collective charge density oscillations
of carriers in the domain wall edge states (Fig.~\ref{Fig1}a), DWPs can
exist even at zero bulk charge density (no doping) with a tunable
frequency from the terahertz up to the mid-infrared ($\sim 200$ meV)
(Fig.~\ref{Fig1}b,c) and disperse linearly in contrast to that
expected from conventional 2D bulk plasmons.

\begin{figure}
  \center \includegraphics[width=8cm]{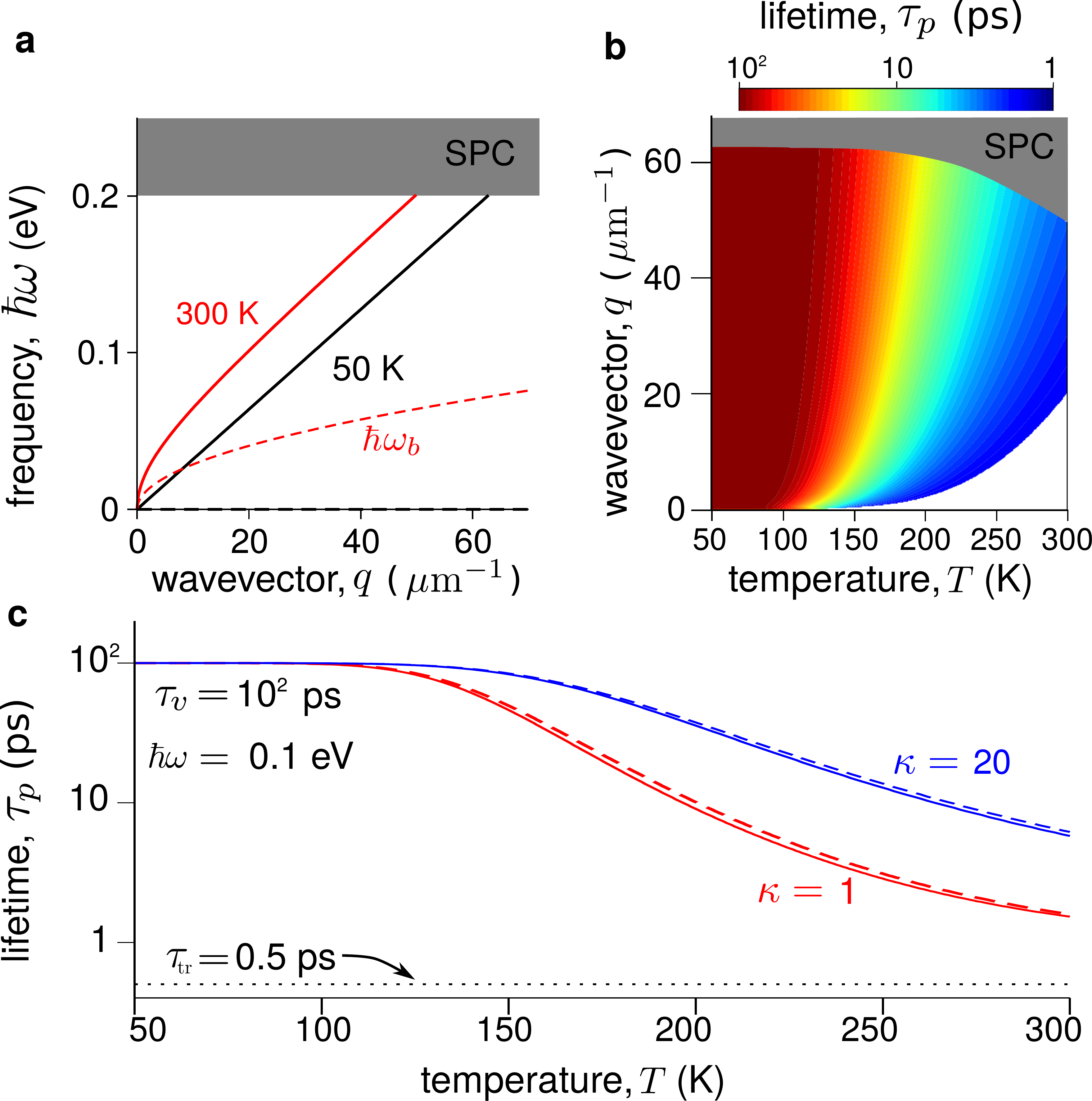}
  \caption{\label{Fig2}{\bf a} DWP dispersion $\omega$ (solid lines)
    for $T = 50 $ K (black) and $T = 300 $ K (red), and bulk plasmon
    dispersion $\omega_b$ (dashed lines). Note that $\omega_b$ for
    $T=50$ K is negligible.  {\bf b} DWP lifetime $\tau_p$ as a
    function of wave vector $q$ and temperature $T$ exhibit large
    values exceeding the transport scattering time by orders of
    magnitude. (top) Gray shaded region indicates the single particle
    continuum (SPC) and white region (right bottom) delineates the
    region where $\omega\lesssim 2.5 \omega_b$. The color bar is in a
    logarithmic scale.  {\bf c} DWP lifetime $\tau_p$ (in log scale)
    as a function of temperature for $\hbar\omega = 0.1\unitev$
    obtained numerically from Eq.~\eqref{eq:disp} (dashed lines) and
    from the estimate in Eq.~\eqref{eq:taup} (solid lines) with
    $\kappa=1$ (red lines) and $\kappa=20$ (blue lines). For
    comparison, we draw bulk transport scattering time
    $\tautr=0.5\unitps$ as a dotted line. We have used parameters
    $\tautr=0.5 \unitps$, $\tau_v=100 \unitps$ and
    $\Delta=0.1\unitev$. }
\end{figure}

Importantly, DWPs are long-lived and possess an insensitivity to bulk
long-range disorder. While conventional plasmon lifetimes are limited
by bulk transport scattering~\cite{soljacic09,sorger12,principi13b},
DWPs at low temperature transcend the restrictions of bulk transport
scattering exhibiting DWP lifetimes orders of magnitude larger than
the bulk transport scattering time (Fig.~\ref{Fig2}).  As we argue
below, these long lifetimes persist to high temperatures and can reach
values of a few picoseconds at room temperature (for corresponding
bulk transport scattering times of $\sim 0.5\, {\rm ps}$).

\begin{figure}[t]
  \center \includegraphics[width=8cm]{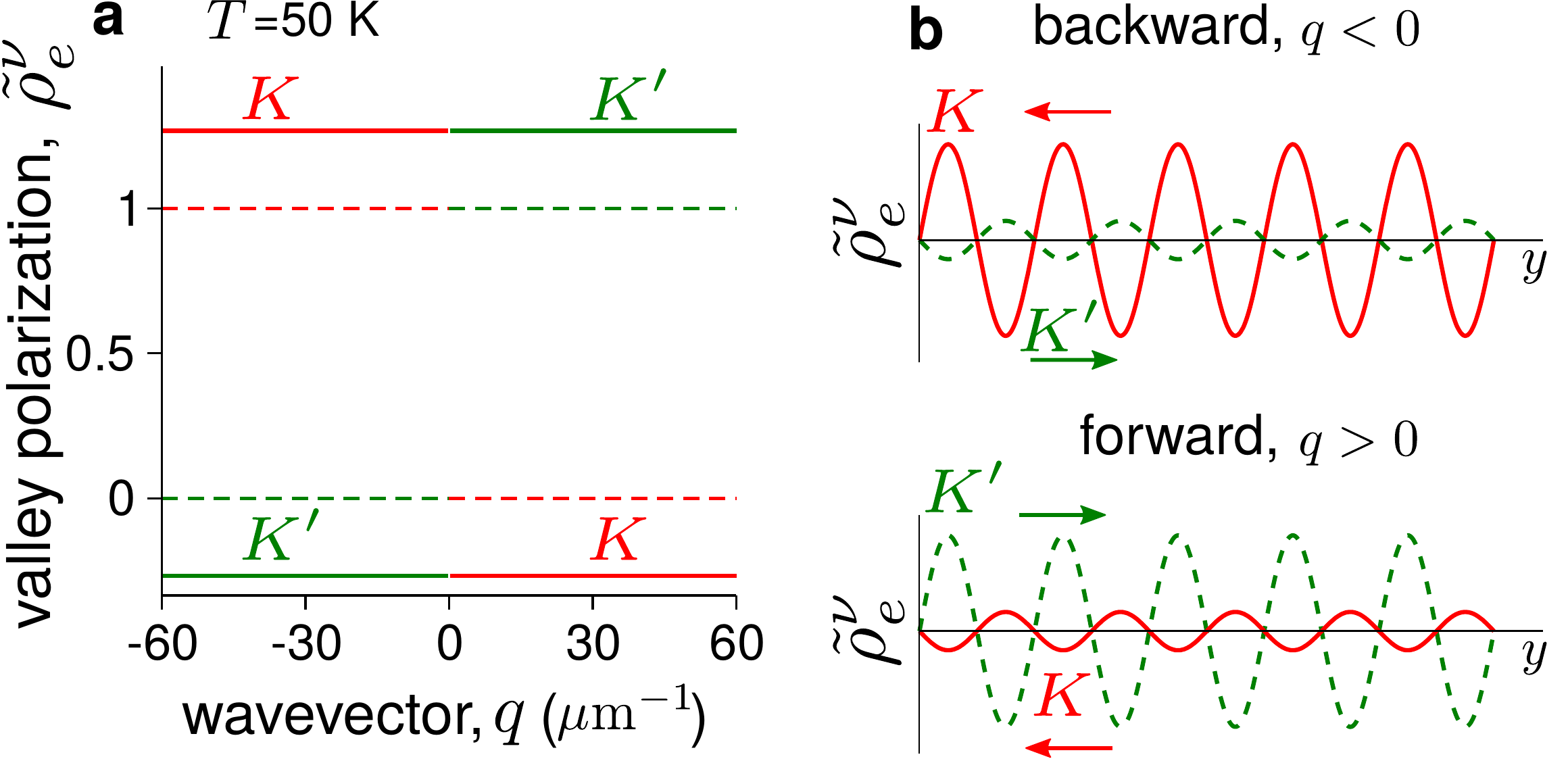}
  \caption{\label{Fig3} Valley polarization
    $\tilde{\rho}_e^\nu=\delta\rho_e^\nu/\delta\rho_e$ where $\nu = K$
    (red solid line) and $K'$ (green solid line) at $T= 50$ K. Dashed
    lines indicate perfect filtering when $\eta \to 0$.  Here we have
    used the parameters $\Delta=0.1\unitev$ and $\kappa=20$. (b)
    Schematic of mixed contribution of $\tilde{\rho}_e^\nu$ for
    forward/backward propagating DWPs.}
\end{figure}

The topological edge states that host DWP are intimately locked to the
difference of valley Chern number on either side of the domain
wall~\cite{martin08,fanzhang13,eunah13}; DWPs possess valley
polarization with backward/forward modes predominantly propagating in
K/K' valleys (Fig. \ref{Fig3}).  As we explain below,
in addition to currents in the domain walls, DWP propagation also
induces bulk undergap valley current flow, which renormalize the
frequency of collective oscillations in the domain wall states.
Control of the latter (e.g, via screening from a dielectric
background) grants an unconventional knob to tune a myriad of DWP
characteristics that range from its velocity and confinement, to the
degree of DWP valley polarization.

We expect DWPs to manifest in experimentally available gapped bilayer
graphene
systems~\cite{fanzhang13,eunah13,martin08,jeiljung11,fengwang14,helin16,junzhu16}
such as along AB/BA stacking faults in globally gapped bilayer
graphene, as well as electrostatically defined domain walls in
split-dual-gate geometries. Indeed, both these methods have been
recently employed to study topological domain walls
experimentally~\cite{fengwang14,helin16,junzhu16}.  DWPs also feature
subwavelength confinement of light, and can be probed by a variety of
techniques that include gratings, and scanning near-field optical
microscopy~\cite{fei12,koppens12}. 

\vspace{2mm}
{\bf Domain wall states and collective dynamics ---} We begin by considering
domain walls in gapped bilayer graphene. These domain walls can be
created in a number of ways, for e.g., (i) defined electrostatically
where split-dual gates in bilayer graphene are biased to yield
adjacent regions with layer potential of opposite
signs~\cite{martin08,fanzhang13,junzhu16}, and (ii) at AB-BA stacking
faults where the bilayer graphene is globally
gapped~\cite{fanzhang13,eunah13,fengwang14,helin16}.

We account for both these types of domain walls phenomenologically by
describing gapped bilayer graphene with a spatially varying band gap:
$\tilde\Delta(x)=\pm 2\Delta$ on either side of $x=0$. Reversing its 
sign at $x=0$, the domain walls at the zero node of $\tilde\Delta(x)$
host DWS (Fig.~\ref{Fig1}a). We note, parenthetically, that the
qualitative form of DWS is insensitive to the specific $\tilde\Delta(x)$ profile used since DWS, arising from band inversion, is locked to its zero nodes.
For electric field
defined domain walls, $\tilde{\Delta}(x)$ directly correlates with the
layer potential difference. For domain walls at stacking faults,
however, the physical band gap (layer potential difference) does not
flip in real space. Instead, the chirality (in each valley) in AB and
BA stacking regions are opposite, leading to opposite signs of valley
specific Berry curvature and Chern number~\cite{martin08,eunah13,fanzhang13}. We absorb
this (chirality) sign into an effective
$\tilde\Delta(x)$. DWS are
valley-helical states located inside the bulk band gap with backward
(forward) moving DWS locked to the valley index $K(K')$
(Fig.~\ref{Fig1}a)~\cite{martin08,eunah13,fanzhang13}.  For each valley, there
are four edge states (DWS) with the same helicity propagating along
$\hat{\vec y}$ stemming from layer and spin degrees of
freedom~\cite{martin08,fanzhang13,eunah13,junzhu16}.

In order to describe the dynamics of carriers in DWS, it is useful
to separate out the density into bulk, $\rho_b (\vec r)$ (in $x<0$ and
$x>0$ regions) as well as edge state density $\rho_e (\vec r)$
(situated at $x=0$) via
\begin{align}
\rho^\nu (\vec r, t) & = \rho_{b,>}^\nu (\vec r, t) \Theta (x) +
\rho_{b,<}^\nu (\vec r, t) \Theta (-x) + \rho_e^\nu (\vec r, t) \delta
(x),\nn \vec j^\nu(\vec r, t) & = \vec j_{b,>}^\nu (\vec r,t) \Theta
(x) + \vec j_{b,<}^\nu (\vec r, t) \Theta (-x) + \vec j_e^\nu (\vec r,
t) \delta (x),\label{eq:bulk-edge}
\end{align}
where $\Theta(x)$ is the Heaviside function, and $\nu = \pm 1$ denote
$K(K')$ valleys.  We note that the edge current $\vec j_e^\nu$ in each of the valleys
arises from the chirality of the edge states: $ \vec j_e^\nu(\vec r) =
-\nu v_0\rho_e^\nu(\vec r) \hat{\vec y}$ where the edge states in
valley $K$ and $K'$ possess effective chiral velocity $-v_0 \hat{\vec
  y}$ and $v_0 \hat{\vec y}$, respectively \cite{supplement}.

Bulk charge density evolves dynamically as
\begin{equation}
  \partial_t\rho^\nu_b(\vec r,t)+\nabla\cdot \vec j^\nu_b(\vec
  r,t)=0,\quad \vec j^\nu_b(\vec r,t) =\boldsymbol{\sigma}^\nu
  [-\nabla\phi(\vec r,t)]
 \label{eq:continuity}
\end{equation}
where $-\nabla\phi(\vec r,t)$ is the electric field, and
$\boldsymbol{\sigma}^\nu$ is the bulk conductivity
tensor. $\boldsymbol{\sigma}^\nu$ contains both diagonal,
$\sigma_{xx}$, as well as off-diagonal components, $\sigma_{xy}^\nu$.
In gapped bilayer graphene, the latter arises from valley Hall
currents~\cite{xiao07,tarucha15,yuanbozhang15} and as we will see below, plays an integral
role in DWP dynamics.  Valley dependent Hall motion is characterized
by the sign of the gap as well as the valley index; here we model $\sigma_{xy}^\nu(x) =
\nu\ {\rm sign (x)} \sigma_H,\ \sigma_H = 4e^2/h$ where the factor $4$
corresponds with the number of DWS in each valley~\footnote{Here we
   have noted that there are two branches of DWS per
   valley/spin~\cite{fanzhang13}.}.

Similarly, the dynamics of the edge charge density can be discerned by applying 
the continuity relation to 
Eq.~\eqref{eq:bulk-edge} and matching $\delta$-functions. We obtain 
\begin{equation}
  \partial_t\rho_e^\nu -\nu v_0 \partial_y \rho_e^\nu+ {\mathcal G}\vec j_{b}^\nu \cdot
  \hat{\vec{x}} =
  -\gamma_v(\delta\rho_e^\nu-\delta\rho_e^{-\nu}),
  \label{eq:Kconstitutive}
\end{equation}
where $ {\mathcal G}\vec j_{b}^\nu=\vec j_{b,>}^\nu\big|_{0^+} - \vec
j_{b,<}^\nu \big|_{0^-}$ and we have used $\partial_x \Theta(\pm x) =
\pm \delta (x)$. While the second term describes dynamics arising from
edge current flow within the DWS, the third term arises from bulk
currents impinging into the DWS. The latter contribution include both
valley Hall $\sigma_{H}$ as well as longitudinal $\sigma_{xx}$
currents. Valley relaxation is accounted for via a phenomenological
inter-valley scattering rate $\gamma_v$. 

Collective modes of the domain wall states emerge as self-sustained
density oscillations of
Eq.~(\ref{eq:bulk-edge}-\ref{eq:Kconstitutive}), and electric
potential obeying
\begin{equation} 
  \phi(\vec r,t) = \int d\vec r' U(\vec r, \vec r') \delta \rho (\vec
  r', t), \quad U(\vec r, \vec r') = \frac{1}{\kappa |\vec r - \vec r'|},
\label{eq:coulomb} 
\end{equation}
in the non-retarded limit. Here $U(\vec r, \vec r')$ is the Coulomb
interaction, and $\delta\rho (\vec r,t)=\rho(\vec r,t)-\rho^{(0)}$
where $\rho^{(0)}$ is the equilibrium charge density. Since the system
is translationally invariant along the edge ($y$ direction), DWPs
propagate as waves of form $\phi(\vec r,t)=\tilde{\phi}_q(x,z) e^{i (q
  y-\omega t)}$ and $\delta \rho(\vec
r,t)=\delta\tilde{\rho}_q(x)\delta(z) e^{i (q y-\omega
  t)}$. Hereafter, we concentrate on the fields $\phi, \delta \rho$ at
$z=0$.

In what follows, we will describe collective modes along the domain
wall compactly in terms of $\phi$, by eliminating $\delta \rho$ from
the dynamical equations. To do so, we first note that charge density
localized on the domain wall, $\delta \rho_{q,e}$, produces a jump in the
electric field as
\begin{equation}
\partial_x\tilde{\phi}_q\zerop-\partial_x\tilde{\phi}_q \zerom=
\left(\partial_xU_q\zerop-\partial_xU_q\zerom\right) \delta
\tilde{\rho}_{q,e}.\label{eq:jump}
\end{equation}
where $U_q(x)=\int dk\ e^{ikx}(q^2+k^2)^{-1/2}/\kappa$ is the
effective one dimensional (1D) Coulomb kernel.  In obtaining
Eq.~(\ref{eq:jump}) we have taken the derivative of
Eq.~(\ref{eq:coulomb}), using the plane-wave forms of $\delta
\rho,\phi$ and Eq.~(\ref{eq:bulk-edge}) above.  Importantly, $\delta
{\rho}_{e}=\delta\rho_{e}^K+\delta \rho_{e}^{K'}$ in
Eq.~\eqref{eq:jump} can be directly related to the electric potential
by inverting Eq.~\eqref{eq:Kconstitutive}:
\begin{equation}
  \delta\rho_{e}=-\frac{ (\mathcal{M}^{K'}+\gamma_v){\mathcal G}\vec
    j_b^{K}\cdot\hat{\vec x} +(\mathcal{M}^{K}+\gamma_v){\mathcal
      G}\vec j_b^{K'}\cdot\hat{\vec x}}{\mathcal
    {M}^K\mathcal{M}^{K'}-\gamma_v^2}, \label{eq:edgen}
\end{equation}
where $\mathcal{M}^\nu =\partial_t+\gamma_v-\nu v_0\partial_y$ is an
operator that acts on $\phi(\vec r,t)$ 
and $\nu=\pm 1$ for $K$ ($K'$) valley; note that $\vec j_b^{K,K'}$ depends on $\phi$ directly through Eq.~\eqref{eq:continuity}.

In addition to continuity of $\tilde\phi_q(x)$ and jump in electric
field discussed above, electric potential of the plasmon, $\phi(\vec
r,t)$, also satisfies Eq.~\eqref{eq:coulomb}; this yields $\phi(\vec
r,t)$ as a solution to a non-local integro-differential problem.
Instead, here we adopt a simplified Coulomb kernel
$\tilde{U}_q(x)=~\frac{1}{\kappa}\int dk\ 2q
e^{ikx}/(2q^2+k^2)$~\footnote{$U_q(x)$ yields highly non-local integro-differential
  equation Eq.~\eqref{eq:coulomb}. We note there are other
  methods to (numerically) analyze integro-differential problem, for example, by the
  Wiener-Hopf  method~\cite{volkov88,katsnelson16} or multipole
  expansion~\cite{xia94,kinaret11}. To illustrate the essential features of DWPs, here we adopt a simplified
  Coulomb kernel $\tilde U_q(x)$ whose Fourier transform matches that
  of $ U_q(x)$ up to leading order in $k/q$.} 
  which captures the essential long wavelength features of
$U_{q}(x)$~\cite{fetter85,zabolotnykh16}.  Using simplified
$\tilde{U}_q(x)$, we find $\tilde\phi_q(x)$ follows
\begin{equation}
  (\partial_x^2-2q^2)\tilde{\phi}_q(x)= \frac{-4 \pi}{\kappa} |q|
  \ \delta \tilde{\rho}_q(x). 
  \label{eq:poisson2}
\end{equation}
Since Eq.~\eqref{eq:poisson2} is local, $\phi_q(x)$ profile can be
obtained in a straight-forward fashion as described below.

We first discuss the dispersive features of DWPs, focussing on the
case $\gamma_v =0$ and Fermi energy inside the gap and $T=0$ so that
no bulk carriers are excited; see below for a detailed discussion of
the role of $\gamma_v$ and $\sigma_{xx}$. This yields $\sigma_{xx}
=0$, $\delta \rho_b=0$ in the bulk, and a solution of Eq.~\eqref{eq:poisson2} as
$\tilde{\phi}_q(x)=\phi_0e^{-\sqrt{2}|qx|}$.  Plugging this 
$\tilde\phi_q(x)$ profile into Eq.~\eqref{eq:jump} and
\eqref{eq:edgen}, we obtain the DWP dispersion (Fig.~\ref{Fig1}c):
\begin{equation}
  \omega =v_0 |q| \sqrt{1+\eta},\quad \eta=4\sqrt 2\pi \sigma_H/v_0
  \kappa\label{eq:plasdisp}.
\end{equation}
The first term inside the square root comes from the velocity of edge
state carriers, whereas $\eta$ captures collective bulk valley Hall
motion that moves along the DWP.  We note that for $\hbar\omega\ge
2\Delta$, DWP enters the single particle continuum (SPC) (shaded
region Fig.~\ref{Fig1}c) where particle-hole excitations damp the
plasmon and destroy its coherence. When $q/q_0< 1$, the SPC boundary,
delineated by $\hbar\omega=2\Delta$, is nearly constant.

Strikingly, bulk valley Hall currents renormalize the collective mode
velocity of DWS in Eq.~\eqref{eq:plasdisp}. Estimating $v_0$ from
Ref.~\cite{martin08}, we obtain $v_0= 4 v_F(\sqrt 2
\Delta/t_1)^{1/2}/3$ at zero Fermi energy, where $v_F= 10^6\ {\rm
  m/s}$ is the monolayer Fermi velocity, $t_1=0.3\ {\rm eV}$ is
the interlayer hopping parameter. Choosing $\Delta=0.1$~eV, $\sigma_H
= 1.5\times10^6\ {\rm m/s}$ and $\kappa=1$, we estimate that the
valley Hall contribution can be 27 times larger than the single
particle edge state contribution (see orange bar vs purple bar in
Fig.~\ref{Fig1}c). As a result, DWP group velocity can be five times
larger than $v_0$.

Low plasmon velocities yield tight confinement of light when the
plasmon is hybridized to form plasmon-polaritons. Indeed, taking
$\Delta=0.1\unitev$ we find a plasmon confinement of about 60
  times smaller than free-space wavelength. For example, for
$\hbar\omega = 0.1\unitev$ [below the single particle continuum (SPC)
where $\hbar\omega \ge 2\Delta$, shaded area, right panel of
Fig.~\ref{Fig1}c], this gives a confinement length as small as
$200\unitnm$ (c.f. free-space wavelength for the same
frequency of $12\ \mu\rm m$). Importantly, since $\eta$ depends strongly
on background $\kappa$, screening can dramatically reduce $\eta$ and
DWP velocity, further enhancing the confinement of DWP (dashed line in
Fig.~\ref{Fig1}c; here $\kappa = 20$). For very large $\kappa$
and small $\Delta = 10\unitmev$, velocities dramatically slow down,
giving a confinement that can be squeezed up to 3 orders of magnitude
shorter than the free-space wavelength.

\vspace{3mm}

{\bf Domain wall plasmon lifetime -- } The dynamics of (thermally
activated) bulk charge as well as inter-valley scattering can
contribute to the decay and damping of DWP.  Employing
Eq.~\eqref{eq:continuity} we find bulk charge dynamics:
$-i\tilde{\omega}\ \delta\tilde{\rho}_{q,b}+
\sigma_{xx}\lp-\partial_x^2+q^2\rp\tilde\phi_q = 0$, where we model
the bulk conductivity via a Drude model:
$\sigma_{xx}=D(\theta)/(\gammatr-i\tilde{\omega}) $,
$\gammatr=1/\tautr$ is the carrier scattering rate and $D(\theta)$ is
the Drude weight~\cite{supplement} that depends on
$\theta=k_BT/\Delta$. Here we have used complex $\tilde{\omega}$ to
capture both plasmon oscillations (${\rm Re}(\tilde{\omega})$ denotes
the plasmon frequency) as well as decay dynamics (${\rm
  Im}(\tilde{\omega})$ denotes its inverse lifetime).  Using these,
$\tilde{\phi}_q(x)$ in the bulk takes the form~\cite{supplement}
\begin{equation}
  \tilde{\phi}_q(x)=\phi_0 e^{-k_0|x|}, \quad k_0=\sqrt{2} |q|
  \lp\frac{\tilde{\omega}^2-\omega_b^2+i\gammatr\tilde{\omega}}
          {\tilde{\omega}^2-2\omega_b^2+i\gammatr\tilde{\omega}}\rp^{1/2}, \label{eq:k0}
\end{equation}
where $\omega_b^2=2\pi D(\theta) |q|/\kappa$ is the bulk plasmon
frequency.  We will first treat the case $ |\tilde{\omega}|^2\gg\omega_b^2$.

Substituting $\delta \tilde{\rho}_{q,e}$ from Eq.~\eqref{eq:edgen}
into Eq.~\eqref{eq:jump} we obtain a complex DWP $\tilde{\omega}$ obeying:
\begin{equation}
   k_0\lb \tilde{\omega}^2+2i\tilde{\omega}\gamma_v-(v_0q)^2+\epsilon
     |q| (i\tilde{\omega}-2\gamma_v) \sigma_{xx} \rb
   =\epsilon \sigma_H v_0 |q|^3,\label{eq:disp}
\end{equation}
where $\epsilon=8\pi/\kappa$.
The plasmon dispersion and lifetime can be discerned from
Eq.~\eqref{eq:disp} by writing
$\tilde{\omega}(q)=\omega(q)-i/\tau_p(q)$, where $\tau_p(q)$ is the
DWP lifetime.  Solving Eq.~\eqref{eq:disp} numerically, we plot the
plasmon frequency ($\omega$) and lifetime ($\tau_p$) respectively in
Figs.~\ref{Fig2}a and b; in these, we have used parameters
$\Delta=0.1\unitev$, and $\kappa = 1$ as well as (disorder-limited)
transport scattering time $\tautr=0.5\unitps$ which corresponds to a
relatively high mobility $50,000\ {\rm cm^2/V s}$ that can be realized
in hBN-encapsulated bilayer graphene
samples~\cite{yuanbozhang15,junzhu16}. Here we use a temperature
independent $\tau_{\rm tr}$. We have also used the intervalley scattering
lifetime $\tau_v=1/\gamma_v=100\unitps$ as estimated 
Ref.~\cite{jeiljung11}.
  
In Fig.~\ref{Fig2}a, we show DWP frequency $\omega$ at low temperature
(black solid line) and room temperature (red solid line) in comparison
with their bulk plasmon frequencies $\omega_b$ (dashed lines). We note
that DWP dispersion remains largely linear and exhibits little
difference between room temperature ($300\,{\rm K}$) vs low
temperature ($50\,{\rm K}$) due to the slow increase of the Drude
weight with temperature~\cite{supplement}. The bulk plasmon frequency
$\omega_b$ is negligible at low temperature. However, $\omega_b$
becomes comparable to $\omega$ at small $q$ and large temperatures.
When $\omega =\alpha \omega_b$, kinematics allow DWPs to rapidly decay
into bulk plasmons, when $\alpha$ is of order unity.  While a detailed
analysis of DWP to bulk plasmon emission is beyond the scope of this
work, we delineate this regime in Fig.~\ref{Fig2}b with regions
$\omega\lesssim \alpha \omega_b$ shown in white. As an illustration,
we set $\alpha = 2.5$. See~\cite{supplement} for a detailed comparison
of $\omega$ to $\omega_b$.

Importantly, as shown in Fig.~\ref{Fig2}(b,c), DWPs [obtained
numerically from Eq.~\eqref{eq:disp}] can exhibit very long lifetimes
$\sim 1.5\unitps$ even at room temperature exceeding reported plasmon
lifetime ($\sim 0.5\unitps$) in hBN-encapsulated
graphene~\cite{woessner15}. Strikingly, $\tau_p$ exceeds the bulk
transport scattering time of $\tautr=0.5\unitps$ (dotted black line,
Fig.~\ref{Fig2}c), and clearly demonstrates how DWP $\tau_p$ can
transcend the conventional limit set by bulk transport
scattering~\cite{woessner15,principi13b}. We note
that expected phonon-limited mobility at room temperature for bilayer
graphene can reach values of
$200,000\unitmobility$~\cite{morozov08,castro10}; mobilities of
$125,000\unitmobility$~\cite{wees11} at room temperature have been
reported in hBN-encapsulated graphene. With those values of mobility,
DWP lifetime may reach $\sim 6\unitps$ at room temperature.

Enhanced lifetimes arises due to a suppression of bulk carrier density
that provides a pathway for DWPs to decay. To illustrate this, we
estimate DWP lifetime $\tau_p$ from Eq.~\eqref{eq:disp} by taking the
limit $\omega_b \ll \omega$ and $ \tau_{v,p,{\rm
    tr}}^{-1}
\ll \omega$. In this limit, $\tau_p$ takes on the simple
form~\cite{supplement}:
\begin{equation}
    \label{eq:taup}
    \frac{1}{\tau_p}=\frac{1}{\tau_v}+\frac{1}{\tautr}\lp\frac{2\omega_b^2}{\omega^2}\rp.
\end{equation}
In obtaining Eq.~\eqref{eq:taup} we have additionally assumed
$\tau_v^{-1} \ll \tau_{\rm tr}^{-1}$. In Fig.~\ref{Fig2}c, we show
$\tau_p$ as a function of temperature for a fixed $\hbar
\omega=0.1\unitev$ obtained from Eq.~\eqref{eq:taup} (solid lines) and
compare with numerical results (dashed lines) showing excellent
agreement. Crucially, Eq.~\eqref{eq:taup} shows explicitly how low
bulk carrier density (encoded in the bulk plasmon frequency,
$\omega_b$) quenches the role of bulk transport scattering in DWP
lifetime. Indeed, neglecting $\tau_v^{-1}$ and for $\omega>\omega_b$,
$\tau_p$ is enhanced by a factor of $\sim \omega^2/2\omega_b^2$ over
$\tau_{\rm tr}$; in this regime, $\tau_p$ scales (approximately)
linearly with $\tautr$. Interestingly, the dependence of $\omega_b$
and $\omega$ on $\kappa$ in Eq.~\eqref{eq:taup} indicate that DWP
lifetime at room temperature can be further boosted by screening as
shown in Fig.~\ref{Fig2}c [$\kappa=1$ (red) and $\kappa=20$ (blue)].

$\tau_p$ exhibits a distinct temperature dependence (see
Eq.~\eqref{eq:taup}, Fig.~\ref{Fig2}c).  At high temperature, since
bulk Drude weight is thermally activated, $\tau_p$ similarly displays
an exponential temperature dependence (Fig.~\ref{Fig2}c) sharply
increasing as temperature drops.  However, at low temperature
$\omega_b$ vanishes (black dashed lines in Fig.~\ref{Fig2}a, see also
Eq.~\eqref{eq:taup}).  As a result, intervalley scattering dominates
DWP lifetime cutting the exponential rise of DWP lifetime
$\tau_p\to\tau_v$ (see Fig.~\ref{Fig2}c).

Due to the valley-helical nature of DWS,
$\tau_v$ can in-principle be very large. Indeed,
Ref.~\cite{jeiljung11} reported that 1D channel is insensitive to
backscattering and long range disorder giving a mean free path of
about $100\ \mu\rm m$ corresponding to $\tau_v$ as high as $100\ \rm
ps$. We note that recent transport experiments along both electric
field and stacking fault domain walls report shorter $\tau_v$ of about
a few hundred $\unitfs$~\cite{fengwang14,junzhu16}. Shorter $\tau_v$ in
electric field domain walls may arise from short-ranged disorder that
can scatter between valleys such as grain
boundaries~\cite{fengwang14}, as well as wide electrostatic profile
used to create $\tilde\Delta(x)$ in electric field defined domain
walls.  In the latter, the electrostatic profile is characterized by a
finite effective width $L_0$ of the domain
wall~\cite{junzhu16}. Although the electronic structure of DWS are
relatively independent of $L_0$ -- domain walls arise whenever the
$\tilde{\Delta}(x)$ flips sign -- broad $L_0$ allow additional
non-chiral (non-topological) states that can mediate scattering
between DWS in separate valleys and consequently reduce
$\tau_v$~\cite{junzhu16}. We note that the typical width used
  by Ref.~\cite{junzhu16} was about $L_0\approx 100\unitnm$.  With reduced
$L_0$ and smooth potential profile, $\tau_v$ may reach long ballistic timescales characteristic of
topological edge states.

\vspace{3mm}

{\bf Valley polarization --} Single-particle carrier
transport within DWS are completely filtered by valley
index~\cite{fanzhang13}: at $K (K')$ valley, carriers in DWS propagate in
the $-y$ ($+y$) direction. In contrast, the collective modes of DWP
experience a mixture of both valley contributions since Coulomb
interactions are long-ranged and do not discriminate between valleys.
In order to quantify how much each valley contributes to the collective
motion of DWP, we analyze Eq.~\eqref{eq:Kconstitutive}. For
simplicity, we specialize to the limit $ \gammatr=0,\gamma_v=0$. Using
$\phi(\vec r, t)$ profiles from Eqs.~\eqref{eq:poisson2} and~\eqref{eq:k0}, we obtain
an oscillating charge density in each valley as 
\begin{equation}
\delta\rho_e^\nu (y,t) =
\frac{1}{\omega+\nu v_0 q}\lb \frac{2 k_0 D(\theta)}{\omega}-\nu
2\sigma_H q\rb\phi_0 e^{i(qy-\omega t)}.
\end{equation}
exhibiting finite amplitudes of charge density for both valleys in
both directions.

The distinction between valley contributions for opposing directions
are particularly clear for very small $T$, where $D(\theta)=0$. In
this limit, we find that valley polarization $\tilde\rho_e^\nu
(y,t)=\delta\rho_e^\nu/\delta\rho_e$ are $\pi$ out-of-phase with each
other (i.e. for every $(q,y,t)$, $\tilde \rho_e^\nu$ have opposite
signs) and have different amplitudes~(Fig.~\ref{Fig3}a,b); departing
from perfect valley polarization regime (dashed lines of
  Fig.~\ref{Fig3}a).  Non-zero amplitude $\tilde \rho_e^\nu$ in both
valleys for DWPs (and partial valley polarization) is a direct
consequence of collective motion of bulk valley Hall currents. Indeed,
DWP frequency renormalization in Eq.~\eqref{eq:plasdisp} originates
from mixing of the two valleys. This contrasts with the $\eta=0$ case
in Eq.~\eqref{eq:plasdisp} where DWPs traveling along $q<0$ ($q>0$)
are fully $K$ ($K'$) valley polarized. In Fig~\ref{Fig3}a, we have
set $\kappa=20$ and obtain about $80\%$ vs $20\%$ mixture of
$|\tilde\rho_e^\nu|$. We note that at smaller $\kappa$,
$\tilde\rho_e^\nu$ will deviate even further from perfect filtering
(dashed lines of Fig.~\ref{Fig3}a) due to a stronger Coulomb potential.
 
DWPs are long-lived and possess decay times that
surpass conventional plasmon decay restrictions wherein plasmon
lifetime is limited by the bulk's transport scattering time. This
property is unusual and stems from the distinct origin of DWPs:
collective oscillations of carriers in the edge
states. Indeed, the edge states enable large quality factors
for DWP oscillations which can range from about $10^2$ (at room
temperature) up to $10^4$ (at low
temperatures)~\cite{supplement}. Surprisingly, DWPs' long lifetime and
high quality manifest without sacrificing subwavelength
electro-magnetic confinement.  A tantalizing prospect for utilizing
DWPs are gate-defining topological domain walls in gapped bilayer
graphene for DWP plasmonic waveguides. Together with high quality and
long-lived DWPs, gate-defined domain walls provide a means for
patterning low-dissipation (and valley polarized) plasmonic
circuits~\cite{ebbesen08,sorger12}.

\vspace{2mm} 

 {\bf Acknowledgements ---} We are grateful for useful conversations with Frank Koppens, Niels Hesp, and Mike Schecter. This work was supported by the Singapore
 National Research Foundation (NRF) under NRF fellowship award
 NRF-NRFF2016-05.

\newpage\null\thispagestyle{empty}

\setcounter{equation}{0}
\renewcommand{\theequation}{S-\arabic{equation}}
\makeatletter
\renewcommand\@biblabel[1]{S#1.}

\section{Supplementary Information for ``Long-lived domain wall plasmons in gapped bilayer graphene'' }

\section*{Sign of edge current } 

Edge
current $\vec j_e^\nu(\vec r,t)$ in Eq.~\eqref{eq:bulk-edge} of
  the main text arises
from single particle motion in the domain wall edge states (DWS). We
determine the direction $\vec j_e^\nu(\vec r,t)$ through bulk-edge
correspondence: valley-helical edge $\vec j_e^\nu(\vec r,t)$ propagates in the same
direction as the bulk undergap valley Hall current $\vec j_b^\nu(\vec r,t)$ close to the edge. 

\begin{figure}[t]
  \center \includegraphics[width=6cm]{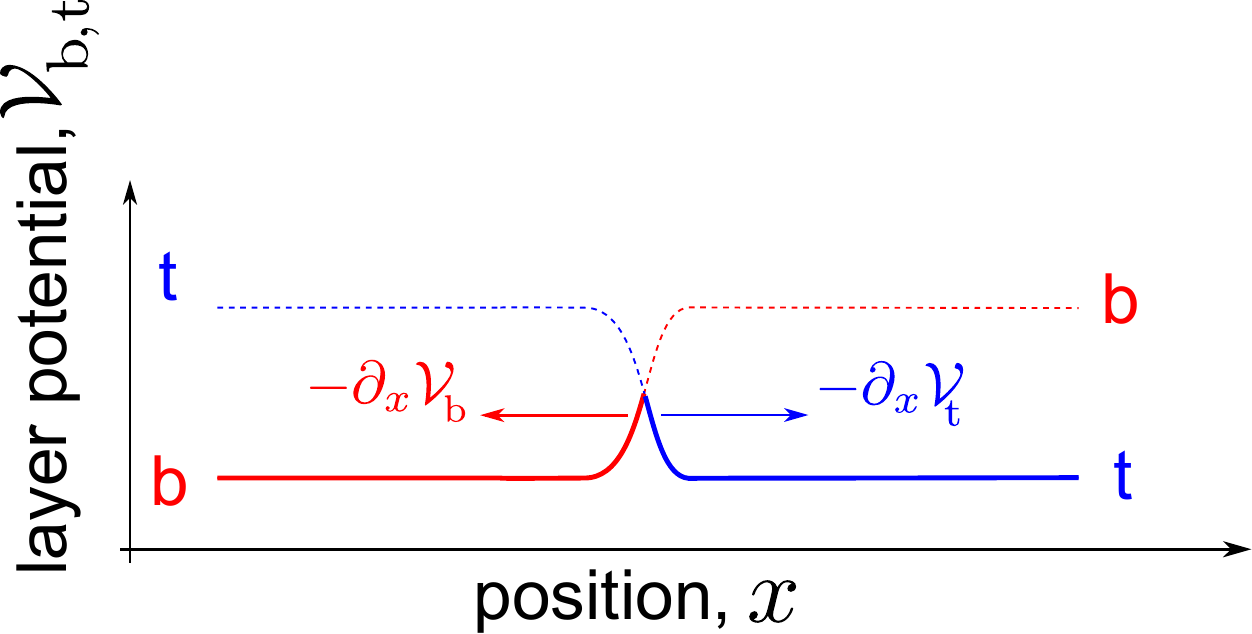}
  \caption{\label{SFig-edge} Sketch of layer dependent potential
    profile for top (t) and bottom (b) layers: ${\cal V}_{\rm b,t}(x)$ (blue line for top layer potential
    ${\cal V}_{\rm t}$ and red line for bottom layer potential ${\cal
      V}_{\rm b}$) flips the sign at the domain wall. Valence (conduction)
    band is highlighted in solid (dashed) lines. Valence band carriers in
    the left (right) side of the domain wall
    reside at the bottom (top) layer experiencing the electric field
    $-\partial_x{\cal V_{\rm b}}\hat{\vec x}$ ($-\partial_x{\cal
      V}_{\rm t}\hat{\vec x}$) directed to $-x$ ($+x$) direction. }
\end{figure}

For clarity, we focus on electric field domain walls where
$\tilde{\Delta}(x)$ reflect the layer potential difference.  We first
note that near an edge, a confining potential $\mathcal V(x)$ creates
electric field $-\partial_x\mathcal V \hat{\vec x}$ and the Hall
current $\vec j_b^\nu(\vec r,t) = -\partial_x\mathcal V \hat{\vec
  x}\times \sigma_{xy}^\nu\hat{\vec z}$. Similarly, close to a domain
wall where $\tilde{\Delta}(x)$ flips sign, a layer dependent potential
$\mathcal V_{\rm b,t}(x)$ exhibits a profile near the domain wall edge
(see solid lines of Fig.~\ref{SFig-edge}), acting
on carriers to produce a bulk $j_b^\nu(\vec r,t)$ in the valence band;
for $k_BT \ll \Delta$ valence band carriers dominate the anomalous
Hall current in each of the valleys. Note that since
$\tilde{\Delta}(x)$ flips sign, valence band carriers reside in
different layers on either side of $x=0$~\cite{martin08,fanzhang13}
and experience different $\mathcal V_{\rm b,t}$ potential profiles.
In $x<0$ ($x>0$) region, $-\partial_x{\mathcal V}_{\rm
    b}\hat{\vec x}$ ($-\partial_x{\mathcal V}_{\rm t}\hat{\vec x}$) is
  pointing along $-x$ ($+x$) as shown in red (blue) arrow. Noting $\sigma_{xy}^\nu(x) = \nu\ {\rm sign (x)} \sigma_H$,
  $\vec j_b^\nu$ in both regions are directed in the $-\nu \hat{\vec
    y}$ direction where $\nu=\pm1$ for $K$ or $K'$, respectively.
Matching the directions of $\vec j_b^\nu$ and $\vec j_e^\nu$, we write $\vec j_e^\nu=-\nu v_0
\rho_e^\nu\hat{\vec y}$.  We expect a similar reasoning also applies for domain
walls at stacking faults where the opposite chirality in AB and BA
  stacking regions flips the sign of effective $\tilde\Delta(x)$.

\section*{Inverse lateral DWP length, $k_0$} 
The self-induced potential around $x=0$ due to DWPs takes the form
$\tilde \phi_q(x)=\phi_0e^{-k_0|x|}$. To obtain the plasmon
inverse lateral 
$k_0$ length in Eq.~\eqref{eq:k0} of the main text
we analyze the dynamics of bulk charge density in
Eq.~\eqref{eq:continuity} of the main text:
\begin{equation}
  -i\omega \delta\tilde\rho_{q,b}(x)+\sigma_{xx}(-\partial_x^2+q^2)\tilde\phi_q(x)=0,
  \label{eq:bulkdyn}
\end{equation}
together with the simplified Coulomb kernel in~Eq.~\eqref{eq:poisson2}
of the main text for $x\ne 0$:
\begin{equation}
  (\partial_x^2-2q^2)\tilde{\phi}_q(x)= \frac{-4 \pi}{\kappa} |q|
  \ \delta \tilde{\rho}_{q,b}(x). 
  \label{eq:poisson2x}
\end{equation}
Substituting $\delta\tilde\rho_{q,b}$ in Eq.~\eqref{eq:poisson2x} with
Eq.~\eqref{eq:bulkdyn} and replacing $\partial_x \to -k_0$, we obtain:
\begin{equation}
  \label{eq:k01}
  (k_0^2-2q^2)\tilde\phi_q(x)=\frac{4\pi}{i\omega\kappa}|q|\sigma_{xx}(k_0^2-q^2)\tilde\phi_q(x).
\end{equation}
Recalling $\sigma_{xx}=D(\theta)/(\gammatr-i\omega)$ and bulk plasmon
frequency $\omega_b^2=2\pi D(\theta)|q|/\kappa$, we can rewrite
Eq.~\eqref{eq:k01} in terms of $\omega_b$:
\begin{equation}
  \label{eq:k02}
 \lb k_0^2\lp1-\frac{2\omega_b^2}{i\omega\gammatr+\omega^2}\rp-2q^2\lp1-\frac{\omega_b^2}{i\omega\gammatr+\omega^2}\rp\rb\tilde\phi_q=0.
\end{equation}
As a result, non-trivial solutions to Eq.~\eqref{eq:k02} occur when 
\begin{equation}
   k_0=\sqrt{2} |q| \lp\frac{\omega^2-\omega_b^2+i\gammatr\omega}
   {\omega^2-2\omega_b^2+i\gammatr\omega}\rp^{1/2}, \label{eq:k0x}
\end{equation}
where we have taken only the positive root to ensure that the
potential profile in Eq.~\eqref{eq:k0} stays finite for all $x$. Note
that in the limit of zero bulk density, $\sigma_{xx} \to 0$, and
Eq.~\eqref{eq:k0x} reduces to $k_0 = \sqrt{2} |q|$.

 \section*{Dynamics of edge charge density}
 Dynamics of charge carriers within the domain wall states along the
 topological domain walls, together with a self-induced electric
 potential, yield DWP collective modes. We note that the dynamics of
 domain wall charge density arise from a number of contributions that
 include the chiral flow of valley charges in each of the edge
 states, bulk undergap valley currents that impinge into the
 edge states, as well as intervalley scattering (which we include via
 a phenomenological intervalley scattering rate
 $\gamma_v=1/\tau_v$). These processes are captured by continuity
 equation of edge charge density as shown in Eq.~\eqref{eq:continuity}
 of the main text, reproduced here for convenience:
 \begin{align}
 \partial_t\rho_e^K -v_0 \partial_y \rho_e^K+ {\mathcal G}\vec j_{b}^K \cdot
 \hat{\vec{x}} =&
 -\gamma_v(\delta\rho_e^K-\delta\rho_e^{K'}),\nonumber \\ \partial_t\rho_e^{K'}
 + v_0 \partial_y \rho_e^{K'} + {\mathcal G}\vec j_{b}^{K'} \cdot
 \hat{\vec{x}} =&
 -\gamma_v(\delta\rho_e^{K'}-\delta\rho_e^{K}),\label{eq:edgK}
 \end{align}
 where ${\mathcal G}\vec j_{b}^K=\vec j_{b,>}^K\big|_{0^+} - \vec j_{b,<}^K
 \big|_{0^-}$. To obtain the charge dynamics in terms of the current dynamics, we invert Eq.~\eqref{eq:edgK}, to find 
  \begin{equation}
    \lp\begin{array}{c} \delta\rho_e^K \\ \delta\rho_e^{K'}
    \end{array}\rp= \frac{-1}{\mathcal {M}^K\mathcal{M}^{K'}-\gamma_v^2}
    \lp\begin{array}{cc} \mathcal{M}^{K'} & \gamma_v\\ \gamma_v &
      \mathcal{M}^K\end{array}\rp \lp\begin{array} {c} {\mathcal G}\vec
      j_b^{K}\cdot \hat{\vec x}\\ {\mathcal G}\vec j_b^{K'}\cdot \hat{\vec x}
    \end{array}\rp,
    \label{eq:invert}
  \end{equation}
  where the operators
  $\mathcal{M}^K=\partial_t+\gamma_v-v_0\partial_y$ and
  $\mathcal{M}^{K'}=\partial_t+\gamma_v+v_0\partial_y$. Summing both
  contributions, the total edge charge density
  $\rho_e=\rho_e^K+\rho_e^{K'}$ is
  \begin{equation}
    \delta\rho_e=-\frac{ (\mathcal{M}^{K'}+\gamma_v){\mathcal G}\vec
      j_b^{K}\cdot\hat{\vec x} +(\mathcal{M}^{K}+\gamma_v){\mathcal
        G}\vec j_b^{K'}\cdot\hat{\vec x}}{\mathcal
      {M}^K\mathcal{M}^{K'}-\gamma_v^2},
     \label{eq:edgensupp}
 \end{equation}
as shown in Eq.~(\ref{eq:edgen}) of the main text.
 
To analyze DWP, we will describe its motion compactly in terms of electric potential, $\phi$, by eliminating $\delta \rho$ from the dynamical equations.
To do so, we first note that the current flow in each of the valleys is directly related to the electric potential via 
Ohm's law [see Eq.~\eqref{eq:continuity} of the main text]. Writing this out explicitly gives
   \begin{align}
     {\mathcal G}\vec j_b^K\cdot\hat{\vec x}&=& -\sigma_{xx}\lp\partial_x\phi^>\zerop -\partial_x\phi^<\zerom
     \rp - 2\sigma_H \partial_y\phi_0,\nn
      {\mathcal G}\vec j_b^{K'}\cdot\hat{\vec x}&=&- \sigma_{xx}\lp\partial_x\phi^>\zerop -\partial_x\phi^<\zerom
     \rp + 2\sigma_H \partial_y\phi_0, \label{eq:gjx}
   \end{align}
 where $\phi_0=\phi\zerox$ [see Eq.~\eqref{eq:k0} of main
     text] and we have noted that the opposite signs of
     $\sigma_{xy}^\nu$ on either side of the domain wall add when
     $\mathcal{G}$ acts on $\vec j_b^\nu$.  Substituting the plane
   wave form 
 $\delta \rho_e(\vec y,t)=\delta\tilde{\rho}_{q,e} e^{i (q
   y-\omega t)}$ into Eq.~(\ref{eq:edgensupp}) and using
 Eq.~(\ref{eq:gjx}), produces a direct relation between $\delta
 \tilde\rho_{q,e}$ and $\phi$:
   \begin{equation}
   \delta \tilde{\rho}_{q, e}
   =\frac{\sigma_{xx}\lp\partial_x\phi^>\zerop -\partial_x\phi^<\zerom
     \rp 2(i\omega-2\gamma_v)+
     4\sigma_Hv_0q^2\phi_0}{\omega^2+2i\omega\gamma_v-(v_0q)^2} ,
   \label{eq:edgerho2}
   \end{equation}

Finally, we recall that $\phi (\vec r,t)$ satisfy boundary conditions
at $x=0$: $\phi(\vec r,t)$ is continuous at $x=0$, $\partial_x \phi$
may exhibit a jump as in Eq.~\eqref{eq:jump} of the main
text. Applying the form of $\phi_q(x)$ in Eq.~\eqref{eq:k0} of the
main text to Eq.~\eqref{eq:edgerho2} and the boundary conditions
above, we obtain the plasmon dispersion (for complex $\tilde{\omega}$) shown in Eq.~\eqref{eq:disp} of the main text. Note that for $\sigma_{xx},\gamma_v \to 0$, Eq.~\eqref{eq:disp} of the main text reduces to Eq.~\eqref{eq:plasdisp} in the main text as expected.

\section*{Drude weight for gapped bilayer graphene} 
The Drude weight for gapped bilayer graphene can be obtained semiclassically via
\begin{equation}
  D = N e^2\int \frac{d^2\vec k}{(2\pi)^2} v^2(k) \lp-\frac{\partial
    f(\epsilon)}{\partial \epsilon}\rp, \label{eq:drudew}
\end{equation}
where $N=4$ accounts for spin and valley degeneracy, $e$ is
electron's charge, $v=\partial \epsilon(\vec k)/ \partial(\hbar \vec k)$ is
electron's group velocity, and $f(\epsilon)$ is the Fermi-Dirac
distribution. We adopt a simple two-band model of gapped bilayer graphene
$\epsilon_\pm(k)= \pm\Delta[1+(k/q_0)^4]^{1/2}$~\cite{fanzhang13,martin08},
where $\Delta$ is half gap,
$q_0=\sqrt{\Delta t_1}/\hbar v_F$, $t_1$ is the interlayer
hopping and $v_F$ is the Fermi velocity of monolayer graphene. 
Using the form of $\epsilon_\pm(k)$ above, we change
integration variables in Eq.~\eqref{eq:drudew} from $k$ to $\epsilon_+$ yielding
\begin{eqnarray}
  D    = \frac{2N
    e^2}{\pi\hbar^2} \int_\Delta^\infty d\epsilon_+
  \frac{\epsilon_+^2-\Delta^2}{\epsilon_+} \lp-\frac{\partial
    f(\epsilon_+)}{\partial \epsilon_+}\rp,
\end{eqnarray}
where the factor of 2 accounts for equal contributions of electrons in the
conduction band and holes in the valence band.
Integrating by parts, recalling $f(\infty)=0$, and making
the integrand dimensionless, $\xi=\epsilon_+/\Delta$, we obtain
\begin{equation}
  D(\theta) = \frac{2N e^2 \Delta}{\pi \hbar^2}{\mathcal F}(\theta), \quad
  {\mathcal F}(\theta)=\int_{1}^{\infty}{d \xi \lp 1+\frac{1}{\xi^2}\rp
    \frac{1}{1+e^{\xi/\theta}}},\label{eq:drudewx}
\end{equation}
where $\theta= k_BT/\Delta$, $k_B$ is the Boltzmann constant and we
have written $f(\epsilon_+)$ explicitly. For all plots in the main text and the supplement, Eq.~\eqref{eq:drudewx} was integrated numerically. 

\section*{DWP lifetime and $\omega/\omega_b$ ratio}
We can estimate DWP lifetime $\tau_p$ from Eq.~\eqref{eq:disp} in the
  main text, reproduced here for convenience:
 \begin{equation} 
   k_0\lb\tilde
   \omega^2+2i\tilde\omega\gamma_v-(v_0q)^2+\epsilon|q|(i\tilde\omega-2\gamma_v)\sigma_{xx}\rb=\epsilon\sigma_Hv_0|q|^3,\label{eq:1}
  \end{equation}
  where $\epsilon=8\pi/\kappa$, $k_0=\sqrt 2 |q|
  [(\tilde\omega^2-\omega_b+i\gammatr\tilde\omega)/(\tilde\omega^2-2\omega_b+i\gammatr\tilde\omega)]^{1/2}$,
  $\omega_b^2=2\pi D|q|/\kappa$
  and $\sigma_{xx}=D/(\gammatr-i\tilde\omega)$. Specializing to the case 
$\omega_b\ll\omega$, yields $k_0=\sqrt 2 |q|$.
Rearranging Eq.~\eqref{eq:1}, we isolate terms containing $\tilde \omega$ to the left hand side
(LHS) so that Eq.~\eqref{eq:1} reads as
\be
f(\tilde\omega)=\frac{\epsilon}{\sqrt 2}\sigma_H v_0 q^2 + (v_0 q)^2,\label{eq:2}
\ee
where 
\be
f(\tilde\omega)= \tilde\omega^2+2i\tilde\omega\gamma_v+\epsilon|q| (i\tilde\omega-2\gamma_v)\frac{D}{\gammatr-i\tilde\omega}. \label{eq:fw}
\ee
It is useful to note that Eq.~\eqref{eq:fw} contains $\tilde{\omega}$ and is a function of complex values, while the RHS is purely real. 
As a result, $\Im f(\tilde\omega)=0$.  
Writing
$\tilde\omega=\omega-i/\tau_p$, this condition can be expressed as
\be
\frac{-2\omega}{\tau_p}+2\gamma_v\omega+\epsilon D|q|\omega\frac{(\gammatr-\gamma_v)}{(\gammatr-\tau_p^{-1})^2+\omega^2}=0. \label{eq:3} 
\ee
Taking the limits, $\gammatr,\tau_p^{-1}\ll \omega$ and 
$\gammatr\gg\gamma_v$, produces a simple relation for the plasmon lifetime as shown in Eq.~\eqref{eq:taup} of the main text, reproduced here for convenience
\be
\frac{1}{\tau_p}=
\frac{1}{\tau_v}+\frac{1}{\tautr} \lp\frac{2\omega_b^2}{\omega^2}\rp\label{eq:lifetime},
\ee
where we have used relations $\gammatr=\tautr^{-1}$ and
$\gamma_v=\tau_v^{-1}$.

\begin{figure}[t]
  \center \includegraphics[width=8cm]{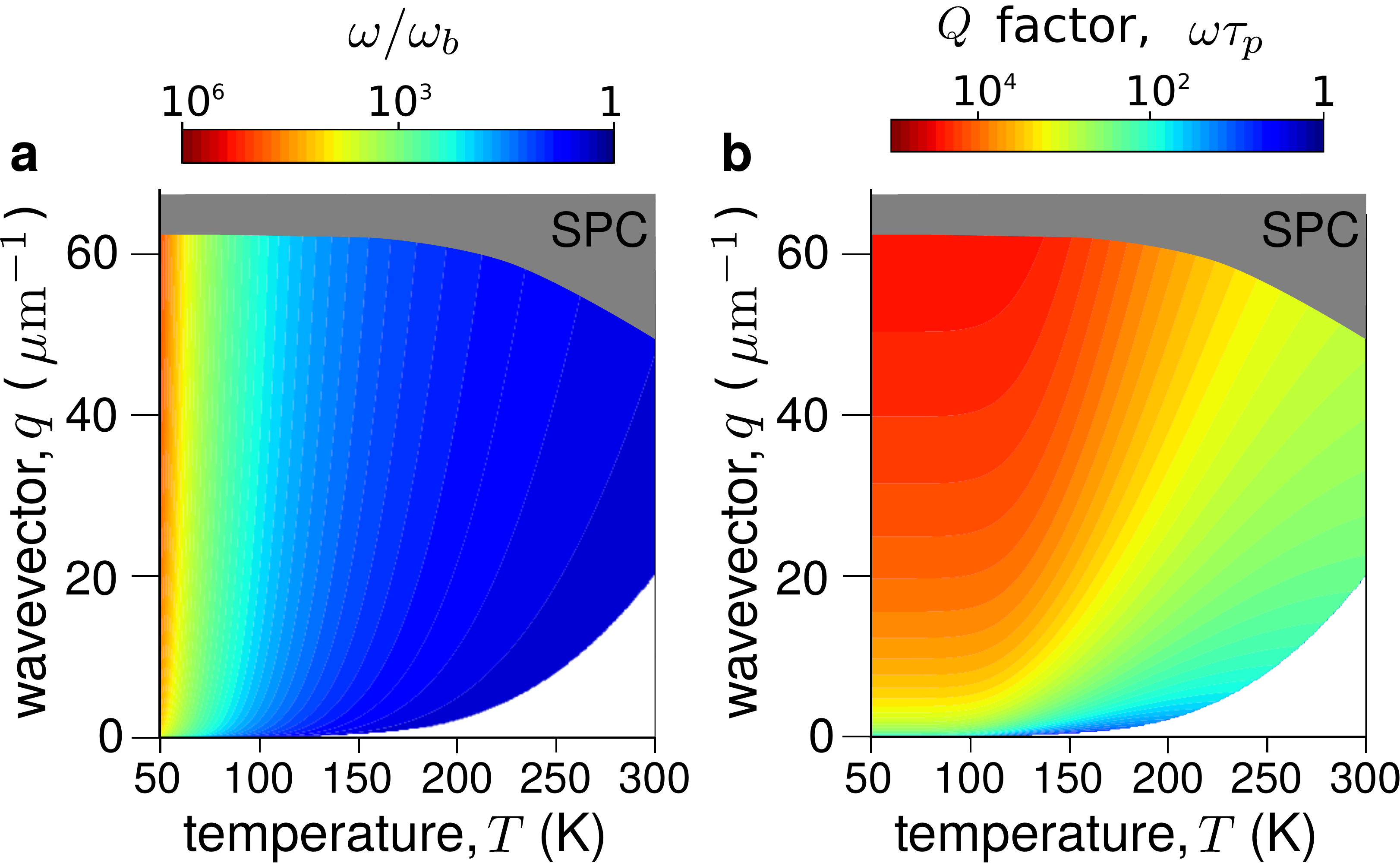}
\caption{\label{SFig3}{\bf a} Ratio of DWP frequency $\omega$ to bulk plasmon 
  frequency $\omega_b$ and {\bf b} $Q$ factor as a function of wave vector
  and temperature.  Shaded region labeled by SPC is the
  single particle continuum (top) and white region (right bottom) 
  indicate $\omega\lesssim 2.5 \omega_b$ regions. The color bars are in
  logarithmic scale. We have used parameters $\tau=0.5 \unitps$,
  $\tau_v=100 \unitps$,  $\kappa=1$ and $\Delta=0.1\unitev$. }
\end{figure}

As shown in Eq.~\eqref{eq:lifetime}, the small ratio between bulk plasmon frequency $\omega_b$ and DWP frequency $\omega$ suppresses the role of $\tau_{\rm tr}$ on DWP lifetime. In Fig.~\ref{SFig3}a, we compare DWP frequency to the frequency of
bulk plasmons $\omega/\omega_b$ as a function of $q$ and $T$ using
Eq.~\eqref{eq:disp} of the main text. Strikingly, at low temperature,
DWP frequency is about six orders of magnitude higher than $\omega_b$
since there are very few carriers in the bulk, resulting a very soft
bulk plasmon mode. Since
$\omega_b\ll\omega$ at low temperatures, decay dynamics from bulk transport scattering is completely quenched. Instead, 
damping from intervalley scattering dominates, 
and $\tau_p\to \tau_v$ [see Eq.~\eqref{eq:lifetime} and
Fig. 2c in the main text].

As temperature increases, the Drude weight in Eq.~\eqref{eq:drudewx}
also increases making the bulk plasmon stiffer. As a result, the
difference between $\omega$ and $\omega_b$ shrinks and the ratio of
$\omega/\omega_b$ drops. Nevertheless, even at room temperature, DWP
can still exceed the bulk plasmon frequency by several times (see red
solid and dashed lines of Fig. 2a in the main text). At high
temperature, scattering from the bulk dominates the $\tau_p$ because
$\gammatr$ is about two orders of magnitude larger than $\gamma_v$.
In Fig. 2c of main text, we displayed that an increase of $\kappa$
prolongs the DWP lifetime when $\omega$ is fixed.  The dependence of
lifetime on $\kappa$ can be discerned from Eq.~\eqref{eq:lifetime} by
approximating $|q|\approx\omega/v_0\sqrt{1+\eta}$. This approximation
was obtained from the wavevector of $T=0$ dispersion for DWP in
Eq.~\eqref{eq:plasdisp} of the main text (we have assumed that $\omega>
\omega_b$). Plugging this estimate into $\omega_b$ of
Eq.~\eqref{eq:lifetime} yields a scaling
\begin{equation}
  \label{eq:taukap}
  \tau_p\approx\tautr\lp\frac{\kappa \omega v_0 \sqrt{1+\eta}}{4\pi D} \rp,
\end{equation}
where we have neglected effect of intervalley scattering at room temperature. 

Recalling $\eta=4\sqrt2 \pi \sigma_H/v_0 \kappa$ and assuming $\eta\gg
1$, we obtain $\tau_p\propto \sqrt \kappa$. Indeed as shown in Fig. 2c
in the main text, $\tau_p$ increases by about $\sqrt {20}$ from
$\tau_p=1.5\unitps$ at $\kappa=1$ to become $\tau_p=6.5\unitps$ at
$\kappa=20$ at $300$ K.

\vspace{3mm}

\section*{Quality factor}
The quality factor or inverse loss function, $Q =
\Re\tilde\omega/\Im\tilde\omega=\omega\tau_p$, is a dimensionless
quantity describing the number of plasmon oscillations performed
before decay.  We plot $Q$ for DWPs in Fig.~\ref{SFig3}b using a
numerical solution of Eq.~\eqref{eq:disp} in the main text showing
large Q factors.  At room temperature, $Q$ for DWPs can be several
hundreds and increases exponentially to about $10^4$ at low
temperatures (Fig.~\ref{SFig3}b). This size of $Q$ is large and
particularly arresting when compared to conventional bulk plasmon $Q$
factors in graphene that have been experimentally observed ($\sim
20$)~\cite{woessner15} and theoretically predicted
($\sim100$)~\cite{soljacic09}.

For ordinary bulk plasmon, dielectrics tend to reduce the $Q$ factor as they introduce
additional scattering pathways for plasmon damping. However, $Q$ of
DWP is surprisingly enhanced by dielectric background at room
temperature and a fixed $\omega$ as $Q\propto \sqrt\kappa$ owing to
Eq.~\eqref{eq:taukap}. The enhanced $\kappa$ reduces the bulk
contribution to DWP which consequently enhances $Q$ factor.


\begin{thebibliography}{10}

\bibitem{kane05}
\bib{Kane, C.~L.; Mele, E.~J.} {Phys. Rev. Lett.} {95} {146802} {2005} 

\bibitem{bernevig06}
\bib{Bernevig, B.~A.; Hughes, T.~L.; Zhang, S.-C.}{Science} {314} {1757--1761} {2006}

\bibitem{konig07}
\bib{ K{\"o}nig, M.; Wiedmann,  S.; Br{\"u}ne,  C.;  Roth, A.;
  Buhmann, H.; Molenkamp,  L.~W.; Qi,  X.-L.; Zhang, S.-C.} {Science}
  {318} {766--770} {2007}

\bibitem{murakami06}
\bib{ Murakami, S.} {Phys. Rev. Lett.} { 97} {236805} {2006}

\bibitem{bernevig06a}
\bib{Bernevig, B.~A.; Zhang,  S.-C.}{Phys. Rev. Lett.} {96} {106802} {2006}

\bibitem{hughes08}
\bib{Qi, X.-L.; Hughes,  T.~L.; Zhang,  S.-C.} {Phys. Rev. B} {78}
  {195424} {2008}

\bibitem{martin08}
\bib{Martin, I.; Blanter,  Y.~M.; Morpurgo, A.~F.} {Phys. Rev. Lett.}
{100} {036804} {2008}

\bibitem{jeiljung11}
\bib{Qiao, Z.; Jung,  J.;  Niu, Q.; MacDonald, A.~H.}  {Nano Lett.}
{11} {3453--3459} {2011}

\bibitem{fanzhang13}
\bib{Zhang, F.; MacDonald,  A.~H.; Mele,  E.~J.} {Proceedings of the National
  Academy of Sciences} {110} {10546--10551} {2013}

\bibitem{fengwang14}
\bib{Ju, L.; Shi,   Z.; Nair,  N.;  Lv, Y.; Jin, C.; Velasco~Jr,  J.; 
  Ojeda-Aristizabal, C.;   Bechtel, H.~A.;  Martin, M.~C.;  Zettl, A.;
  Analytis, J.;  Wang,  F.} {Nature} {520} {650--655} {2015}

\bibitem{helin16}
\bib{ Yin, L.-J.;  Jiang, H.; Qiao, J.-B.; He, L.}  {Nature
  Communications} {7} {11760} {2016}

\bibitem{junzhu16}
\bib{Li, J.;  Wang, K.;  McFaul, K.~J.;  Zern, Z.;  Ren, Y.; Watanabe, K.; 
  Taniguchi, T.; Qiao, Z.;  Zhu, J.} {Nat Nano} {11} {1060--1065} {2016}
  
  \bibitem{eunah13}
\bib{Vaezi, A.; Liang,  Y.; Ngai, D.~H.; Yang, L.;  Kim, E.-A.}
 {Phys.  Rev. X} {3}{021018} {2013}
  
  
\bibitem{castro07}
\bib{Castro, E.~V.; Novoselov,  K.~S.; Morozov, S.~V.; Peres, N.~M.~R.;  dos
  Santos, J.~M. B.~L.; Nilsson, J.; Guinea,  F.;  Geim, A.~K.;  Neto,
  A.~H.~C.}
{Phys. Rev.  Lett.} {99} {216802} {2007}

  
  \bibitem{soljacic09}
\bib{Jablan, M.; Buljan, H.; Solja\ifmmode \check{c}\else
  \v{c}\fi{}i\ifmmode~\acute{c}\else \'{c}\fi{},  M.} {Phys. Rev. B}
{80} {245435} {2009}

\bibitem{sorger12}
\bib{ Sorger, V.~J.; Oulton, R.~F.; Ma, R.-M.;  Zhang, X.}{MRS Bulletin}
  {37} {728–738} {2012}

\bibitem{principi13b}
\bib{Principi, A.; Vignale,  G.; Carrega, M.;  Polini, M.} {Phys.
  Rev. B} {88} {121405} {2013}
 
\bibitem{fei12}
\bib{Fei, Z.; Rodin, A.~S.; Andreev, G.~O.; Bao, W.; McLeod, A.~S.;~Wagner, M.; 
  Zhang, L.~M.;~Zhao, Z.;~Thiemens, M.;~Dominguez,  G.; Fogler, M.~M.;
  Neto,   A.~H.~C.;  Lau, C.~N.;~Keilmann,  F.;  Basov, D.~N.}
{Nature} {487} { 82--85} {2012}

\bibitem{koppens12}
\bib{Chen, J.; Badioli, M.; Alonso-Gonzalez, P.; Thongrattanasiri, S.;
  Huth, F.;  Osmond, J.; Spasenovic,  M.;  Centeno, A.;  Pesquera, A.;
   Godignon, P.;  Zurutuza~Elorza, A.;  Camara, N.;~Garcia
  de~Abajo, F.~J.;  Hillenbrand, R.;  Koppens, F. H.~L.}{Nature} {487} {77--81} {2012}



\bibitem{supplement}
See {\bf Supplementary Information} for a discussion of the sign of edge current, inverse lateral DWP length $k_0$, dynamics of
  edge charge density and lifetime, Drude weight for gapped bilayer graphene, and $\omega/\omega_b$ ratio and Q 
  factor.

\bibitem{xiao07}
\bib{~Xiao, D.; Yao, W.; Niu, Q.} {Phys. Rev. Lett.} {99} {236809}
{2007}

\bibitem{tarucha15}
\bib{~Shimazaki, Y.; ~Yamamoto, M.;  Borzenets, I.~V.; Watanabe, K.; Taniguchi, T.;~Tarucha,   S.}{Nat Phys} {11} {1032--1036} {2015}

\bibitem{yuanbozhang15}
\bib{ Sui, M.; Chen, G.;  Ma, L.;  Shan, W.-Y.;  Tian, D.;  Watanabe,
K.;   Taniguchi, T.;  Jin, X.;  Yao, W.; Xiao, D.;  Zhang,  Y.}{Nat
  Phys} {11} {1027--1031} {2015}

\bibitem{fetter85}
\bib{ Fetter, A.~L.}{Phys. Rev. B} {32}{7676--7684}{1985}

\bibitem{zabolotnykh16}
\bib{ Zabolotnykh, A.~A.; Volkov,  V.~A.}{JETP Letters} {104} {411--416} {2016}

\bibitem{woessner15}
\bib{ Woessner, A.; Lundeberg, M.~B.;  Gao, Y.;  Principi, A.; 
  Alonso-González, P.;  Carrega, M.;  Watanabe, K.;  Taniguchi, T.; 
  Vignale, G.;  Polini, M.;  Hone, J.;  Hillenbrand, R.; 
  Koppens, F. H.~L.}{Nat Mater} {14} {421--425} {2015}
  
\bibitem{morozov08}
\bib{ Morozov, S.~V.;  Novoselov, K.~S.;  Katsnelson, M.~I.; Schedin,
  F.;  Elias, D.~C.;  Jaszczak,  J.~A.;  Geim,
  A.~K.}{Phys. Rev. Lett.} {100} 
{016602} {2008}

\bibitem{castro10}
\bib{ Castro, E.~V.; Ochoa, H.;  Katsnelson, M.~I.;  Gorbachev, R.~V.;
   Elias, D.~C.;   Novoselov, K.~S.;  Geim, A.~K.; Guinea, F.}
 {Phys. Rev. Lett.} {105} {266601} {2010}

\bibitem{wees11}
\bib{ Zomer, P.~J.;  Dash, S.~P.; Tombros, N.;   van Wees, B.~J.}{Applied Physics
  Letters} {99} {232104} {2011}


\bibitem{ebbesen08}
\bib{Ebbesen, T.} {Physics Today} { 5} {44} {2008}

\bibitem{volkov88}
\bib{ Volkov, V.~A.;  Mikhailov, S.~A.}{Zh. Eksp. Teor. Fiz.} {94} {217--241}
  {1988}


\bibitem{xia94}
\bib{Xia, X.; Quinn,  J.~J.}{Phys. Rev. B} {50} {8032--8034} {1994}

\bibitem{kinaret11}
\bib{ Wang, W.;  Apell, P.;  Kinaret, J.}{Phys. Rev. B} {84} {085423} {2011}
  
\bibitem{katsnelson16}
\bib{Principi, A.;  Katsnelson, M.~I.;  Vignale, G.}{Phys. Rev.
  Lett.} {117} {196803} {2016}

\end{thebibliography}
\end{document}